\documentclass[12pt]{iopart}

\usepackage{color}

\usepackage{todonotes}

\begin{document}


\title{Testing gravity with cold atom interferometry: Results and prospects}

\author{Guglielmo M. Tino}

\address{Dipartimento di Fisica e Astronomia and LENS Laboratory\\
Universit\`a di Firenze and INFN-Sezione di Firenze\\ via  Sansone 1,  Sesto Fiorentino, Italy}
\ead{guglielmo.tino@unifi.it}
\vspace{10pt}
\begin{indented}
\item[] 15/12/2020
\end{indented}

\begin{abstract}


Atom interferometers have been developed in the last three decades as new powerful tools to investigate gravity. 
They were used for 
measuring the gravity acceleration, 
the gravity gradient, 
and the gravity-field curvature, 
for the determination of the gravitational constant, 
for the investigation of gravity at microscopic distances, 
to test the equivalence principle of general relativity and the theories of modified gravity,
to probe the interplay between gravitational and quantum physics and to test quantum gravity models,
to search for dark matter and dark energy,    
and they were  proposed as new detectors for the observation of gravitational waves.
Here I describe past and ongoing  experiments with an outlook on what I think are the main prospects in this field and the potential to search for new physics.

\end{abstract}

%
%
%
%
%

%
\tableofcontents
\eject

%
%
\section{Introduction}
\label{sec:Intro}
%
%


Atom interferometers \cite{Berman1997,Cronin2009,Tino2014}, as well as atomic clocks \cite{Poli2013,Ludlow2015}, are powerful tools for precision measurements and fundamental tests in physics \cite{Safronova2018}  and for applications \cite{Bongs2019}.

In this paper, I focus on experiments using atom interferometers to investigate gravity for fundamental physics tests. I describe past and ongoing  experiments with an outlook on what I think are the main prospects in this field and the potential to search for new physics.

The discussion of the experiments is organized in the different sections according to their main motivation but it is worth emphasizing that the same experiment  can have different interpretations and physical implications depending on the results and on the underlying theoretical model.

%
%
\section{Measuring gravity with atoms}
\label{sec:MeasGrav}
%
%

The first demonstration of a measurement of gravity acceleration using  cold atom interferometry was published about thirty years ago in \cite{Kasevich1991} and later with a higher precision in \cite{Peters1999}.

The workings of atom interferometers can be  understood by analogy with  optical interferometers: using  atom optics tools made of material structures or, nowadays more often, laser light, the   wave packet of the atoms entering the system is split, reflected and recombined: at the output, an interference signal can be observed if no which-way information is available. 
More generally,  atom interferometry can be considered as an example of quantum interference due to the different paths connecting the initial and the final state of a system. 
Any physical effect, such as gravity, acting in a different way for the different paths will lead to a change of the interference pattern; by measuring this change, the effect can be studied. 

Recent experiments are usually based on atom interferometry schemes in which  the wavepackets of freely falling cold atoms are split and recombined with laser pulses \cite{Kleinert2015};  two-photon Raman \cite{Kasevich1991,Kasevich1992,Peters1999}
or Bragg \cite{Giltner1995,Muller2008b,Altin2013,Damico2016} transitions or single-photon transitions on ultranarrow lines \cite{Hu2017,Rudolph2020} are used to prevent spontaneous emission processes.
The gravity acceleration $g$ produces a phase change 
at the interferometer output 
\begin{equation}
\Delta \phi \propto k g T^2~,
\label{phase}
\end{equation}
where $k$ is the effective wavevector of the light  splitting and recombining the wavepacket and $T$ is the free-fall time for the atom between the laser pulses. This corresponds  to the free-fall distance measured with the laser wavelength as a ruler.

Other schemes were demonstrated to measure $g$ with atom interferometry as, for example, in experiments based on Bloch oscillations~\cite{BenDahan1996,Clade2005,Ferrari2006}. In the case of Bloch oscillations, the cold atoms are held in a vertical optical lattice; the  effect of gravity and of the periodical potential due to the laser standing wave produces oscillations in momentum space with a frequency $\nu_{BO}$ given by
%
\begin{equation}
\nu_{BO}=\frac{mg \lambda}{2h}~,
\label{BO}
\end{equation}
where $m$ is the atomic mass, $\lambda$ is the wavelength of the laser producing the lattice and $h$ is Planck's constant. By measuring the frequency of the Bloch oscillations, the gravity acceleration $g$ can be determined.
This can be interpreted as the measurement of the difference in the gravitational potential between adjacent lattice wells which are separated by $\lambda/2$. 
Since just a few wells must be filled with ultracold atoms to observe the Bloch oscillations,  this gravimeter can have a sub-millimiter size down to a few micrometers. For this reason, it was also proposed and developed as a method to test  gravity at micrometric distances \cite{Ferrari2006,Sorrentino2009}.

Different schemes, based on Raman, Bragg, and Bloch, can be  combined to increase the interferometer performances \cite{Mueller2009,Charriere2012,Zhang2016,Xu2019}.
Atom interferometry using magnetic pulses instead of light pulses was also demonstrated \cite{Amit2019,Margalit2020}.

Using atoms as quantum probes to investigate gravity is interesting by itself and offers different advantages compared to macroscopic masses. 
The most important is that new experiments are possible taking advantage of the  specific features of atomic sensors:  tests can be performed with  masses having well-defined properties such as proton and neutron number, spin, internal quantum state, bosonic or fermionic nature.
For precision measurements, possible systematics can be drastically reduced due to
 the well known and reproducible properties of the atoms, the 
small size and precise control of the position of atomic samples, the potential immunity from stray field effects, and the possibility of using
different states and different isotopes to reject spurious effects and
 cross-check the results.

Several physical effects were investigated using atom interferometers. 
In particular, as described in the following,  atom interferometry can be used in gravitational physics for
measuring the gravity acceleration~\cite{Kasevich1992, Peters1999, Ferrari2006, Muller2008, LeGouet2008,Hu2013, Sorrentino2012, Freier2016,Hardman2016,Xu2019, DAmico2019,Heine2020}, 
the gravity gradient~\cite{Snadden1998, Mcguirk2002,Sorrentino2012,Sorrentino2014, Duan2014, Pereira2015, Wang2016,Damico2016}
and the gravity-field curvature~\cite{Rosi2015, Asenbaum2017}, 
for the determination of the gravitational constant $G$
\cite{Mcguirk2002, Tino2002,Fattori2003, Bertoldi2006, Fixler2007, Lamporesi2008, Rosi2014, Prevedelli2014}, 
for the investigation of gravity at microscopic distances \cite{Ferrari2006,Sorrentino2009,Tackmann2011},
to search for dark matter \cite{Arvanitaki2015,Geraci2016}, dark energy,  chameleon and test theories of modified gravity \cite{Hamilton2015,Jaffe2017,Sabulsky2019}.
Atom interferometry was used to test the weak equivalence principle of general relativity \cite{Tino2020} by comparing the free fall of different atoms, $^{85}$Rb vs $^{87}$Rb~\cite{Fray2004,Bonnin2013, Zhou2015,Asenbaum2020}, $^{39}$K vs $^{87}$Rb~\cite{Schlippert2014}, the bosonic $^{88}$Sr vs the fermionic $^{87}$Sr~\cite{Tarallo2014}, atoms with different spin orientations~\cite{Tarallo2014,Duan2016}. 
 Tests of the weak equivalence principle using  atom interferometers in space were proposed~\cite{Tino2007_space,Tino2013,Altschul2015}.
Experiments on anti-hydrogen are in progress~\cite{Kellerbauer2008,Charman2013}.
%
%
Atom interferometers, also in combination with  optical atomic clocks, were  proposed for the observation of gravitational waves  \cite{Chiao2004,Aspen2004,Roura2006,Tino2007,Dimopoulos2008,Tino2011,Yu2011,Hogan2011,Graham2013,Hogan2016,Kolkowitz2016,Tino2019} and the first prototypes are presently under construction \cite{Canuel2018, Graham2017,Adamson2018,Zhan2019}.

As can be seen from Eq.~\ref{phase}, the sensitivity of an atom interferometer as a gravimeter increases with the square of the interrogation time and with the effective wavevector of the light. This motivated the development of  atom-optical elements based on multi-photon momentum transfer~\cite{Muller2009, Chiow2011,Mazzoni2015,Rudolph2020} and of large-scale facilities providing a few seconds of free fall time~\cite{Dimopoulos2007,Zhou2011,Kovachy2015}. Eventually it will lead to experiments in space \cite{Tino2007_space,Tino2013,Altschul2015,Tino2019} for which the technology development is in progress~\cite{Aguilera2014}, and proof-of-principle experiments were recently performed~\cite{Becker2018}. 

Long interferometer times with freely-falling atoms  require atomic samples with temperatures in the pK range that can be achieved using ultracold atom sources  and collimation methods \cite{Chu1986,Ammann1997,Morinaga1999,Muntinga2013,Kovachy2015}.
An alternative approach is using coherent matter-wave guides, either optical~\cite{Ferrari2006,Poli2011,Xu2019,Nelson2020} or magnetic~\cite{Berrada2013,Pandey2019}, which  can enable  interrogation times of several seconds in  compact devices.

For a given momentum transfer and interrogation time, the interferometer sensitivity is limited by the so-called quantum projection noise. Work is in progress to overcome this limit and potentially reach the Heisenberg limit by introducing quantum correlations between the individual atoms thus producing squeezed atomic states \cite{Cox2016,Hosten2016,Salvi2018,Pezze2018,Szigeti2020}.

%

In addition to fundamental physics, atomic gravimeters and gravity gradiometers can be used for applications in geophysics and geodesy \cite{Bongs2019} on ground~\cite{DeAngelis2009,deAngelis2011,Schmidt2011,Freier2016,Chiow2016,Menoret2018,Wu2019}, and for Earth observation and planetology in future space missions \cite{Bresson2006,Douch2018,Migliaccio2019,Trimeche2019,Leveque2019,Abrykosov2019}. These applications of atom interferometry will not be discussed here.

%
\section{Determination of the gravitational constant $G$}
\label{sec:MeasG}
%

The Newtonian constant of gravity $G$ is a fundamental physical constant which has been measured in several experiments for long time but it is still the one  known with the lowest precision because of large discrepancies among the values obtained in different experiments. Although this is most probably due to uncontrolled systematics and underestimated errors, the possibility of yet hidden physical effects cannot be excluded.  An overview of the experimental efforts and open problems to determine the value of $G$ can be found in the theme issue {\it ``The Newtonian constant of gravitation, a constant too difficult to measure?''} \cite{TheoMurphy2013}. 

$G$ appears in the well-known equation for the gravitational force between two masses:
\begin{equation}
{\bf F}( {\bf r}) = - G \; \frac{m_1 m_2}{r^2} \bf\hat{r}~.
\label{force}
\end{equation}
The  weakness of the gravitational interaction and the impossibility of shielding the effects of gravity make it very difficult to measure $G$ with high precision keeping systematic effects  under control. 
The difficulty in getting a precise number for $G$ is then paradigmatic of how difficult it is to measure gravitational effects precisely.

    Since  there is no complete theory linking gravity to the other forces of nature, there is no definite relationship between $G$ and the other fundamental constants and no theoretical prediction for the value of $G$ against which testing the experimental results. 

    Despite the numerous measurements performed since the historical experiment  by Cavendish in 1798 \cite{Cavendish1798}, the uncertainty on $G$ has improved only by less than three orders of magnitude in about two centuries. In fact, even the results of the most precise measurements reported by different groups  show substantial discrepancies by parts in $10^4$ between each other so that in the 
    %
    %
    2018 CODATA recommended values of the fundamental physical constants, the value of the Newtonian constant of gravity is 
$G$~=~6.67430(15)$\times 10^{-11}$~m$^3$kg$^{-1}$s$^{-2}$ 
with a relative uncertainty of 2.2$\times 10^{-5}$. 
    
    The realization of conceptually different experiments is then important to try and identify the origin of the discrepancies and  improve the confidence in the final result. 

    Most of the experiments performed so far, including recent ones \cite{Armstrong2003,Luo2009,Quinn2013,Newman2014,Li2018}, were based on the torsion pendulum or torsion balance scheme as in the experiment  by Cavendish. Some experiments were based on different schemes: a beam-balance system \cite{Schlamminger2006}, a laser interferometry measurement of the acceleration of a freely falling test mass \cite{Schwarz1998}, experiments based on Fabry-Perot or microwave cavities \cite{Ni1999, Kleinevoss1999,Parks2010}. They were all based however on the use of macroscopic masses for the gravity source and for the probe.
%
    
    In \cite{Rosi2014,Prevedelli2014}, for the first time $G$ was measured with high precision using an atom interferometer as the probe. 
%
%
The basic idea of the experiment was to use an atom interferometer as  gravity sensor and a well-characterized mass as the source of a gravitational field. From the precise measurement of the gravity acceleration produced by the source mass and from the knowledge of the mass distribution, the value of the gravitational constant was determined.
%
%
%
    A detailed description of the development of the experiment, named MAGIA as the acronym for ``Accurate Measurement of $G$ by Atom Interferometry'',  can be found in \cite{Tino_inVarenna}. It was initially proposed in 2001 and the construction of the apparatus started in 2002 \cite{tino01bis,Tino2002,Fattori2003};  preliminary results were published in \cite{Bertoldi2006,Lamporesi2008}. Proof-of-principle results of another conceptually similar experiment \cite{Mcguirk2002} were published in \cite{Fixler2007}.
%
%
    The MAGIA apparatus was designed with the specific aim of the accurate determination of $G$. 
      The challenge was not only  reaching a high sensitivity in the detection of the gravitational effect produced by the source mass but mostly in the control and reduction of possible systematic effects. This idea guided the design of the atom interferometry sensor and the source mass configuration. 
    Efforts were therefore devoted to the control of systematic effects related to atomic trajectories, positioning of source masses, and stray fields. 
Raman atom interferometry was used to perform precision measurements of the differential acceleration experienced by two samples of laser-cooled $^{87}\textrm{Rb}$ atoms in a vertical gravity gradiometer configuration \cite{Sorrentino2010,Sorrentino2014}
under the influence of nearby  source masses. 
The source mass was made of $\approx \! 500~\textrm{kg}$ of tungsten in two sets of cylinders positioned around the vertical magnetically shielded interferometer tube  \cite{Lamporesi2007}. During the experiment, they were moved in different positions in order to modulate the relevant effect and perform a differential detection. 
The distance of the source mass from the atoms was kept large enough, at the expense of the signal size, in order to reduce the sensitivity to the horizontal size of the atomic cloud  that would produce a systematic effect. Also, we found a configuration taking advantage of the high density of tungsten to compensate the Earth's gravity gradient thus reducing the sensitivity to the vertical position and size of the atomic clouds.
The double differential configuration  drastically reduced numerous common-mode spurious effects. The measurement was modeled by a numerical simulation taking into account the mass distribution and the evolution of atomic trajectories. The comparison of measured and simulated data provided the value of the Newtonian gravitational constant $G$.    
 The result of the MAGIA experiment was $G=6.67191(99) \times 10^{-11}$~m$^3$kg$^{-1}$s$^{-2}$ with a relative uncertainty $\Delta G/G = 1.5\times 10^{-4}$.
This is to date the most precise measurement of $G$ obtained with atom interferometry and it was included in the CODATA adjustment of the recommended values of the fundamental constants of physics \cite{Mohr2016}. The experiment also allowed us to identify limits of the apparatus and showed possible directions for improvements. 
The main limits in the accuracy were indeed the non-negligible atomic velocity distribution and the knowledge of the source mass distribution.

In \cite{Tino_inVarenna,Rosi2014}, ideas for a higher precision measurement of $G$ were mentioned based on the following key features:
a highly homogeneous source mass,
a high-sensitivity atom interferometer,
a better definition of atomic velocities and a smaller size of the atomic sensor,
a scheme to  determine accurately the distance of the atomic source from the source mass,
atoms with a small sensitivity to magnetic fields.
	As far as the source mass is concerned, a possibility would be to use gold which has a high density and  is known to have a high homogeneity if properly  processed; of course,  the source mass should be much smaller than the one used in our experiment. The ``perfect'' source mass would eventually be  silicon  that can be produced as defect-free, ultra-pure monocrystalline samples whose internal structure is extremely regular and can be accurately characterized; the density in this case is about one order of magnitude smaller compared to  tungsten and gold. A higher sensitivity atomic probe would then be necessary in both cases.  The need of a smaller atomic probe suggests to use ultracold or Bose-Einstein condensed atoms confined in an optical lattice. The lattice would also allow to set the atoms at a very precise  distance  from the source mass. Atoms insensitive to magnetic fields would simplify the experiment,  avoiding for example the need of Zeeman pumping, and could be brought close to the source mass.
	Our  experience to date suggests the choice of Sr atoms and silicon source mass  to satisfy the requirements listed above. 
Strontium has indeed a special combination of features:
we showed that precise measurements of gravity can be performed with Sr atoms confined in optical lattices \cite{Ferrari2006,Poli2011,Zhang2016}; the possibility of  efficient and fast cooling of Sr down to BEC was  demonstrated \cite{Stellmer2013b}; we showed in ref.~\cite{Sorrentino2009} how Sr atoms can be positioned at a very well defined distance from a source mass using an optical lattice; $^{88}$Sr has an extremely small collisional cross-section and no magnetic moment in the ground state, making it a perfect atomic probe.
	Combining an atomic sensor based on ultracold Sr atoms in an optical lattice with a high-homogeneity silicon source mass makes it possible to envisage the possibility of reaching a precision in the ppm range for the measurement of $G$.

Different new experiments are in progress or planned to determine $G$ using atom interferometry sensors.

In an experiment  in Stanford,   the apparatus  is based on a horizontal gravity gradiometer atom interferometer with the source masses placed between the two sensors \cite{Biedermann2015}. This scheme with the symmetric source mass configuration is expected to reduce the sensitivity to atom-source positioning. Lead bricks were used as the source mass. The results showed the possibility of reaching a precision $\Delta G/G \approx 10^{-4}$ with the prospect of further improvement.

An experiment conceptually similar to MAGIA  was started in Wuhan \cite{Duan2014}. The sensor is a vertical gravity gradiometer based on Raman interferometry with Rb atoms.  Stainless steel spheres  symmetrically placed around the vertical interferometer tube will be used as sources masses. The planned precision is $\Delta G/G \approx 10^{-4}$.

In Florence, a new experiment  started recently. The scheme was proposed in \cite{Rosi2017_Metrologia}; it is based on an atomic vertical gravity gradiometer as in the previous experiment but, thanks to colder atomic probes, an improved design of the source masses, and the implementation of a method for the cancellation of the gravity gradient phase shift~\cite{Roura2017,DAmico2017},  the systematic effects due to the cloud size, temperature and trajectories will be  reduced. The goal is to reach a precision $\Delta G/G \approx 10^{-5}$ and beyond.

The possibility of a measurement of $G$ based on the gravitational Aharonov-Bohm effect  \cite{Hohensee2012PRL} was discussed in \cite{Mueller2014_inVarenna}. In this case, the measurement is not based on the force but on the gravitational potential difference between saddle points. This idea is interesting by itself and has potential advantages in terms of precision: since at the saddle points the potential is constant up to quadratic terms, errors due to the uncertainty of the relative position of the source masses and the atoms can be reduced. Also, small source masses can be used that can be made of highly homogeneous materials. 


The construction of the apparatus for an experiment based on the ideas outlined in  \cite{Tino_inVarenna,Rosi2014} started recently at  Northwestern University \cite{Kovachy2019_Priv_Comm}. The plan is to use evaporatively cooled lensed clouds of strontium atoms and Bragg large-momentum-transfer atom interferometers with macroscopic scale delocalizations and single crystal silicon proof masses that are horizontally alternated between near and far configurations. The final goal is a measurement of $G$ at 10 ppm  level or better.

As a prospect, it can be expected that in the next decade precision measurements of $G$ with atomic sensors will lead to a better understanding of yet hidden systematics. 
They will also provide tests of possible new physics as for example,  a possible  deviation from the $1/r^2$ law that would lead to a dependence of the value of $G$ on the source-probe distance that was already taken into account in the analysis of some previous measurements \cite{Quinn2001}.  
An interesting topic is also the possible space-time dependence of the value of $G$ which is predicted by gravitational theories alternative to general relativity. It should be noted however that $\dot{G}/G$  is strongly constrained by astronomical observations at the level of $\sim 10^{-13}$ yr$^{-1}$ \cite{Will2014}.

%

%
\section{Testing the $1/r^2$ Newtonian law and gravity at small distances}
\label{sec:TestNewt}
%

Testing the $1/r^2$ Newtonian law and the investigation of gravity at small spatial  scales  is an important challenge for present research in physics in the search for deviations from Newtonian gravity due to physics beyond the standard model, new boson-exchange forces, extra space-time dimensions, possible connection with the small observed size of Einstein cosmological  constant and as  tests of general relativity \cite{Arkani1998,Arkani1999,Hoyle2001,Adelberger2003}.  

Possible deviations from Newtonian gravity  are usually described assuming a Yukawa-type potential
    \begin{equation}
    V(r)=-G \;\frac{m_1m_2}{r}(1+\alpha e^{-r/\lambda}) \label{eq:r2}~,
    \end{equation}
where $G$ is Newton's gravitational constant, $m_1$ and $m_2$ are the masses, $r$ is the distance between them, the parameter $\alpha$ gives the relative strength of departures from Newtonian gravity, and $\lambda$ is its spatial range.
 
Most experiments  searching for deviations at small distances used  as a sensor a torsion pendulum~\cite{Adelberger2009,Sushkov2011} or a microcantilever~\cite{Geraci2008}.
    Experiments with torsion pendula and microcantilevers have set bounds  for $\alpha$ down to micrometer spatial scales. Recent results showed that any gravitational-strength $(|\alpha| =1)$ Yukawa interaction must have $\lambda$ smaller than $\sim 40 ~\mu$m \cite{Lee2020,Tan2020}. At shorter ranges the experimental limits are less stringent.


    The small size and the high sensitivity of  atomic sensors may enable a direct, model-independent measurement at sub-mm distances down to a few $\mu$m from the source mass with no need for modeling and extrapolation as in the case of macroscopic probes. This would allow us to  access directly  regions in the $\alpha-\lambda$ plane which are still unexplored. 
%
%
 Using atom interferometry for the investigation of gravity at micrometric distances was  proposed in refs.~\cite{tino01bis,Tino2002,Dimopoulos2003,Ferrari2006}. The possibility of using atoms to study  effects close to a surface, such as Casimir effect, was also investigated in~\cite{Carusotto2005,Wolf2007,Beaufils2011}.
Preliminary results using atoms as a probe were reported in~\cite{Harber2005,Obrecht2007a,Obrecht2007b} by detecting perturbations of the frequency of the center-of-mass oscillations of a trapped atomic Bose-Einstein condensate near a surface.

Early experiments with Sr atoms on optical frequency references using visible intercombination lines \cite{Tino1992,Ferrari2003} and towards Bose-Einstein condensation \cite{Sorrentino2006} showed us that strontium is a good choice not only for optical clocks but also for atom interferometry. In particular,  the $^{88}$Sr isotope in its ground state can be an ideal probe for precision gravity measurements, even at small distances, because of its extremely small collisional cross section \cite{Ferrari2006a,Stein2010}  and insensitivity to external perturbations due to its null magnetic moment. 
This was first proposed and demonstrated in \cite{Ferrari2006} by observing persistent Bloch oscillations of the atoms in a vertical optical lattice.  Bloch oscillations with high visibility for $\sim 20$~s were later reported in \cite{Poli2011} and methods to increase the precision in the measurement of the Bloch oscillation frequency using lattice modulations to induce tunneling between neighboring sites of the vertical optical lattice were demonstrated in \cite{Ivanov2008,Alberti2010}.

All the tools required for an experiment on gravity at micrometric distances were demonstrated in \cite{Sorrentino2009}: the accurate positioning of the atoms at a few micrometers from a surface was obtained   by applying a relative frequency offset to the counterpropagating laser beams producing the lattice thus translating the atomic sample in a controlled way.
For experiments at distances below $10~\mu$m, the atomic sample size was  compressed  using an optical tweezer. 
   	In order to subtract non-gravitational effects, such as Casimir and Van der Waals forces, the source mass was covered with a gold conductive screen.
The gold coating acted as a mirror to produce the optical standing wave and as a conductive screen. Common-mode effects would be subtracted by performing differential measurements with different source masses  behind the shield \cite{Bennett2019}. 
In a first experiment combining the atom elevator and the lattice modulation method to probe effects close to a glass surface, 
 a  broadening of the resonance and a reduction of the signal was observed when the atoms were brought to a distance $< 1$~mm from the surface \cite{Tino_inVarenna}. Tests performed on different glass samples showed  similar results with a dependence of the size of the effect on the glass surface. The observed effect can be attributed  to stray light from the glass sample due to spurious reflections or scattering producing speckles  that affect the optical lattice and lead to decoherence. A related effect might have been observed in ref.~\cite{Charriere2012}. The observed spurious effect is a limitation for the use of Bloch oscillations to investigate gravity at sub-mm distances from a source mass
because  the sensitivity of the atomic probe is dramatically  reduced. Further work would be required to understand the origin of the effect and to control it.

A different method to induce coherent tunneling between neighboring sites of a vertical 1D optical lattice separated by the Bloch frequency using Raman laser pulses was demonstrated with $^{87}$Rb in \cite{Beaufils2011, Tackmann2011} and proposed as a scheme to measure short-range forces \cite{Wolf2007}. Preliminary results showing a spatial resolution of $ 3~ \mu$m with a sensitivity of $5 \times 10^{-6}$ at 1~s for the measurement of the Bloch frequency was reported in \cite{Alauze2018}. The experiment is in progress, the next main step being the transport of the atoms close to a surface and the test of the method for the measurement of forces at sub-millimeter/micrometer distances.

A new strontium atom interferometry experiment that aims to probe the gravitational inverse square law at length scales of $0.1-1$~m is under construction \cite{Kovachy2020_Priv_Comm}.  Previous work has identified atom interferometry as a promising candidate to study this range of length scales \cite{Biedermann2015}.  This project plans to leverage large momentum transfer atom optics, single-crystal silicon proof masses, and a combination of ultracold atoms and spatially resolved atom detection to improve sensitivity and reduce systematic errors.  Following the method of Ref.~\cite{Asenbaum2017}, large momentum transfer atom optics will be used to split an initial atom cloud into two separated atom clouds in order to form a vertically-oriented gravity gradiometer that measures the gravitational signal from a local proof mass.  To reduce systematic errors arising from uncertainty in the atomic trajectories, an ultracold atom source \cite{Kovachy2015} and spatially resolved atom detection \cite{Dickerson2013} will be implemented.  As atom interferometric gravitational measurements are further improved, systematic errors arising from density inhomogeneities in the proof mass will become increasingly important.  As mentioned above, a promising approach to ameliorate this effect is to use single-crystal silicon, which is highly homogeneous \cite{Tino_inVarenna,Rosi2014}.  Atom interferometric gravitational measurements with single-crystal silicon proof masses have not yet been practical because of the comparatively low density of silicon which makes it more difficult to measure a gravitational signal with sufficient resolution.  To overcome this limitation, ultrasensitive atom interferometers employing large momentum transfer  atom optics will be employed.  The construction of the atom interferometry apparatus as well as the design of the proof mass system are currently in progress.  Ultimately, the experiment aims to probe values of the  coupling strength $\alpha$ for a Yukawa-type  force down to the level of $10^{-5}$ in the 0.1~-~1~meter length scale range. This experimental setup will also be used for a new measurement of Newton's gravitational constant (see Sect.~\ref{sec:MeasG}).

The possibility of an experiment with a Cs gravity gradiometer to set constrains for $\lambda \sim 10$~cm was discussed in \cite{Biedermann2015}. Preliminary results showed that $\alpha$ near $10^{-5}$ could be reached with an  improvement of about two orders of magnitude over existing limits. 

The schemes developed to measure gravity produced by small source masses with the main motivation of investigating dark energy \cite{Jaffe2017,Xu2019,Sabulsky2019} might also be used for the investigation of short-range forces.

%
%
\section{Experimental tests of the weak equivalence principle \label{sec:TestsWEP}}
%
%

The equivalence principle is the basis of general relativity. Testing it  corresponds then to testing the validity of general relativity \cite{Will2014}. Its weak form, the weak equivalence principle of general relativity, namely, the universality of free fall that corresponds to the equivalence of the gravitational and inertial mass, was verified  to a remarkable accuracy with different kinds of experiments. A recent wide review of the theoretical background and implications of the equivalence principle and of the experimental tests can be found in Ref.~\cite{Tino2020} from which most of the contents of this section derive.  

In general, the experiments testing the weak equivalence principle look for a  small differential acceleration $|a_1-a_2|$ between two freely falling test masses of different nature. Possible violations of the weak equivalence principle are expressed in terms of the  E\"{o}tv\"{o}s parameter $\eta$:
\begin{equation}
\eta=2\left |\frac{a_1-a_2}{a_1+a_2}\right |.
\end{equation}

Different kinds of  experiments were performed to test the weak equivalence principle. 
On Earth, torsion balances provided so far the best bounds on possible  violations with a relative precision of $\sim 10^{-13}$ \cite{Schlamminger2008,Adelberger2009,Zhu2018}. 
In space, the MICROSCOPE mission provided the most accurate test of the weak equivalence principle with a relative precision of about $10^{-14}$ \cite{Touboul2017,Touboul_2019}. Stringent bounds were set also by lunar laser ranging measurements \cite{Merkowitz2010,Hofmann2018}. It is worth noting that other experiments rely on the weak equivalence principle for their validity. Examples are the measurement of the gravitational constant $G$  with freely falling samples \cite{Schwarz1998,Rosi2014} and the comparison of different gravimeters  \cite{Francis2015}.

Here, the tests performed using atom interferometry and their prospects are discussed. The possibility of testing the equivalence principle using atom interferometry has been indeed the main motivation for several experiments and for ongoing efforts to develop ever more sophisticated apparatus. 



%


%

Compared to experiments with macroscopic masses,  the main interest of using atoms is 
that qualitatively new tests  can be performed thanks to their quantum features by comparing atoms with different properties like proton and neutron number, spin, internal quantum state, bosonic or fermionic nature.

%



In Ref.~\cite{Peters1999}, the Raman interferometry gravimeter using Cs atoms was compared with a classical gravimeter based on a freely falling corner-cube. The results showed that  the macroscopic glass mirror falls with the same acceleration as the Cs atoms to within 7 parts in $10^{9}$. More recently,  mobile Raman atom gravimeters with $^{87}$Rb and classical absolute gravimeters with similar uncertainties were compared for metrological purposes  \cite{Merlet2010,Francis2015}. 
In \cite{Poli2011}, the value of gravity acceleration measured with a gravimeter based on Bloch oscillations of Sr atoms in a vertical optical lattice was compared with the value measured in the same lab with a classical  gravimeter. The two values agreed within 140 parts in $10^{9}$.


The weak equivalence principle was tested in experiments with different isotopes of an atomic species. The similar masses and transition frequencies make the setups and the control of systematics less complex compared to  experiments with different atoms. 
Several experiments were performed with the two isotopes of rubidium, $^{85}$Rb and $^{87}$Rb, mainly because  the required experimental tools have been developed for many years.
Gravity acceleration for $^{85}$Rb and $^{87}$Rb was first compared in \cite{Fray2004} with a relative accuracy of $\sim 10^{-7}$ using an atom interferometer based on the diffraction of atoms from standing optical waves. A test for a possible dependence of the free fall acceleration from the relative orientation of nuclear and electron spin was also performed with $^{85}$Rb atoms in two different hyperfine states. 
A similar precision was later obtained in \cite{Bonnin2013} using Raman atom interferometry for the differential free fall  of $^{85}$Rb and $^{87}$Rb.
A four-wave double-diffraction Raman interferometry scheme was used in \cite{Zhou2015} to compare gravity acceleration for $^{85}$Rb and $^{87}$Rb in a simultaneous dual-species atom interferometer. The value obtained for the E\"otv\"os parameter was $\eta = (2.8 \pm 3.0 \times 10^{-8})$. The optimization of the apparatus and a new test  at a level of precision  $\sim 10^{-10}$ were reported in \cite{Duan2020,Zhou2019}.
Ongoing experiments with Rb in large-scale interferometers are aiming to a precision of $10^{-15}$  and beyond \cite{Dimopoulos2007,Zhou2011,Kovachy2015}. 
Possible limits due to the gravity gradients were discussed in \cite{Nobili2016} and a solution was proposed in \cite{Roura2017} and  demonstrated in \cite{DAmico2017,Overstreet2018}. 
In \cite{Overstreet2018}, thanks to the compensation of the gravity gradient in a long-duration and large-momentum-transfer dual-species interferometer, a relative precision of $\Delta g/g \approx 6 \times 10^{-11}/$shot or $3 \times 10^{-10}/\sqrt{Hz}$ was demonstrated showing the feasibility of this test at the $10^{-14}$ level. Recently, a dual-species atom interferometer based on a large-momentum-transfer sequence of Bragg transitions with 2s of free-fall time was used to measure the relative acceleration between $^{85}$Rb and $^{87}$Rb at the level of  $10^{-12}g$ \cite{Asenbaum2020}; this is the best result obtained so far with atomic sensors approaching the $10^{-13}$ precision limit of the methods based on macroscopic probes in Earth laboratories.
 The weak equivalence principle was tested also for the $^{88}$Sr and $^{87}$Sr isotopes of strontium \cite{Tarallo2014}. Gravity acceleration was measured from the frequency of the Bloch oscillations for the two isotopes in a vertical optical lattice. The value obtained for the E\"otv\"os parameter was $\eta = (0.2 \pm 1.6 \times 10^{-7})$. As discussed in the following, the results reported in \cite{Tarallo2014} have relevance also to the tests of the equivalence principle for bosons vs fermions and for the investigation of spin-gravity coupling.

%
%
Tests of the weak equivalence principle with different atoms were started. The interest of experiments with different atoms and their sensitivity to  violations of the equivalence principle predicted in a dilaton model and in extensions of the standard-model were discussed in \cite{Damour2012,Hohensee2013b,Hartwig2015}.  
More complex experimental apparatus and a more difficult control of  systematics are required in these experiments compared to tests with different isotopes of the same element so the precision achieved until now is lower. 
A test with rubidium and potassium atoms was discussed in \cite{Varoquaux2009} and the first results were reported in \cite{Schlippert2014,Barrett2016,Albers2020}. In \cite{Albers2020},  $^{39}$K and $^{87}$Rb   were compared using two Raman interferometers; the result for the E\"otv\"os ratio was $\eta = (-1.9 \pm 3.2 \times 10^{-7})$.  
The work in progress towards a test with rubidium and ytterbium atoms in a 10-m baseline atom interferometer was discussed in \cite{Hartwig2015} with the prospect to reach a precision in the $10^{-12}-10^{-13}$ range. 
In Florence, a new atom interferometry apparatus is under construction that will enable experiments with strontium and cadmium atoms \cite{Poli2018_ICAP,Tinsley2019_ECAMP}.

While some of the tests with atoms described above might be considered as analogous to the ones performed with macroscopic classical objects, experiments have been proposed and performed in which the quantum features of the atoms as probes of gravity are essential.



%

In \cite{Zych2015,Zych2018},  a quantum test of the equivalence principle  was proposed based on the idea that, because of the mass-energy relation $E=mc^2$ of special relativity, the internal energy of a system affects its mass. In addition to the interest of testing the equivalence principle for atoms in different energy eigenstates, of particular importance is in this frame a test with atoms in  superpositions of the internal energy states because this corresponds indeed to a genuine quantum test. A related test was proposed in \cite{Orlando2016}.
The first  test of the equivalence principle in the quantum formulation was reported in \cite{Rosi2017}. Using Bragg atom interferometry in a gravity gradiometer configuration, the gravity acceleration values for $^{87}$Rb atoms  in two hyperfine states $|1\rangle= |F=1, m_F=0\rangle$ and $|2\rangle=|F=2, m_F=0\rangle$, and in the coherent superposition $|s\rangle=(|1\rangle + e^{i\gamma}|2\rangle)/\sqrt{2}$ were compared. 
An upper bound of $5 \times 10^{-8}$ was obtained for the parameter corresponding to a  violation of the weak equivalence principle for a quantum superposition state. 
For atoms in the $|1\rangle$ and $|2\rangle$ hyperfine states, an E\"otv\"os ratio $\eta_{1-2} = (1.0 \pm 1.4 \times 10^{-9})$ was obtained corresponding to an improvement by about two orders of magnitude with respect to the previous limit reported in \cite{Fray2004}.
A further improvement on the latter test was reported in \cite{Zhang2018,Zhou2019}, approaching the $10^{-10}$ level. Based on models \cite{Damour2012} in which the violations of the equivalence principle are expected to increase with the separation in energy  between the internal levels, in \cite{Rosi2017} an experiment involving states with a larger energy separation was  proposed as an interesting prospect: in particular, optically separated levels in strontium were considered for which atom interferometry was already demonstrated \cite{Mazzoni2015,Hu2017,Hu2020}. 

%

%

Other tests of the weak equivalence principle with atoms were proposed for which quantum physics is crucial.
 
A test of the weak equivalence principle for atoms in entangled states was proposed in \cite{Geiger2018} and preliminary results towards a possible experiment were reported in \cite{Shin2019}. 
%
%

 In \cite{Viola1997}, the free fall of particles in  Schr\"odinger cat states in  configuration space was investigated theoretically. 

A possible difference in the gravitational interaction for fermions and bosons was discussed in \cite{Barrow2004} and a first  experiment with $^{87}$Sr and $^{88}$Sr was reported in \cite{Tarallo2014}.

 A difference in the  free fall for different atoms in Bose-Einstein condensates is envisaged in models considering spacetime fluctuations and the extended wavepackets (\cite{Goklu2008,Herrmann2012} and references therein). The fluctuations would also lead to decoherence.
The search for these effects requires high-sensitivity atom interferometry and a  long evolution time so a prospect is to perform such experiments in microgravity \cite{vanZoest2010,Debs2011,Kuhn2014,Altschul2015,Sondag2016,Becker2018}.

%


%

Possible spin-gravity coupling and torsion of space-time were investigated theoretically \cite{Peres1978,Ni2010,Capozziello2011} and the effects were searched for using macroscopic test masses \cite{Ni2010,Heckel2008,Zhu2018}, atomic magnetometers \cite{Venema1992,Kimball2013}, and in the hyperfine resonances in trapped ions \cite{Wineland1991}. The  free-fall experiments with atoms in different hyperfine states are also important in this frame \cite{Fray2004,Rosi2017,Zhang2018}.
Using Bloch oscillations in a vertical optical lattice, in \cite{Tarallo2014} gravity acceleration was measured for the bosonic $^{88}$Sr, which has zero total spin in its ground state, and for the fermionic  $^{87}$Sr, which has a half-integer nuclear spin $I=9/2$. An E\"otv\"os parameter $(0.2 \pm 1.6) \times 10^{-7}$ was obtained. The analysis of the Bloch resonance spectrum for $^{87}$Sr including the different Zeeman states, allowed to set an upper limit for the coupling of spin to gravity and for the neutron anomalous acceleration and spin-gravity coupling \cite{Venema1992,Ni2010}.
%
%
In \cite{Duan2016}, gravity acceleration for $^{87}$Rb in different Zeeman states  was compared  using a Raman atom interferometer. 
The resulting E\"otv\"os parameter was $(0.2 \pm 1.6 )\times 10^{-7}$. 
%
In \cite{Rodewald2018},  the prospect of testing the weak equivalence principle for molecules with opposite chiralities was mentioned.

The production of low-energy antihydrogen atoms \cite{Amoretti2002,Gabrielse2002} opened the way to precision tests of the weak equivalence principle for neutral antimatter \cite{Walz2004,Kellerbauer2008,Hamilton2014,Perez2015}.
Comparing the gravitational properties of matter and antimatter allows to test standard model extensions \cite{Kostelecky2011} and quantum vacuum \cite{Hajdukovic2010}. 
%
%
Early experiments to test gravity for electrically charged particles and antiparticles \cite{Witteborn1967,Holzscheiter1995} were generally limited by stray electric and magnetic field effects \cite{Schiff1970,Darling1992}. 
 A preliminary measurement of the Earth's gravitational effect on magnetically trapped antihydrogen provided an upper bound of 100 times
 $g$ \cite{Amole2013}. Current efforts are mainly devoted to increasing the rate of production of antihydrogen and reducing the temperature in order to enable precision spectroscopy and gravity measurements using atom interferometry.
Tests of the weak equivalence principle for antimatter could also be performed  with muonium \cite{Kirch2014,Antognini2018} and with positronium \cite{Cassidy2014,Sala2019}.

In conclusion, atom interferometry  enabled  precision tests of the weak equivalence principle  that were previously performed only with macroscopic classical masses.
The sensitivity of atomic experiments did not reach yet that of  classical experiments but it can be anticipated that a similar or  higher  precision will be obtained. Perhaps more important is that qualitatively new tests of the weak equivalence principle can be devised with  atom interferometry taking advantage of the quantum nature of atomic gravity sensors. In the future,  matter-wave interferometry with molecules might enable  tests for systems with  different conformations,  internal states,  chiralities \cite{Rodewald2018}. The tests of the weak equivalence principle  were performed so far only for systems consisting of particles of the first elementary particle family while direct tests for particles of the second and third family are missing (\cite{Antognini2018} and references therein).

The effort to increase the sensitivity for the equivalence principle tests pushed the development of atom interferometers based on atomic fountains with several meters baseline \cite{Zhou2011,Dickerson2013} and others are being developed in Hannover, Berkeley, Florence. The final precision would be reached in experiments in  space as in the proposed STE-QUEST \cite{Aguilera2014,Altschul2015} and SAGE \cite{Tino2019} missions (see Sect.~\ref{sec:largescale}).

%

\section{Probing the interplay between gravity and quantum mechanics}

Our present understanding of physical phenomena is based on two different theories which are incompatible with each other: quantum mechanics describes correctly the microscopic world of atoms, molecules, elementary particles while general relativity  describes gravity and the large-scale behavior of the universe. We do not have a quantum theory of gravity.

In this section, experiments performed or proposed with the specific goal of probing the interplay between gravity and quantum mechanics using atom interferometry are discussed. Section \ref{sec:DMAI} is devoted to the discussion of related work in the frame of quantum gravity.

 The results of experiments measuring $g$ with atom interferometry~\cite{Peters1999,Clade2005, Ferrari2006,Ivanov2008} were reinterpreted in~\cite{Muller2010a} as measurements of the gravitational redshift. Precise measurements of the gravitational redshift are important to verify the local position invariance and test the Einstein equivalence principle.
The basic concept of the analysis in \cite{Muller2010a} is to consider the Compton frequency $\omega_C = m c^2/\hbar$ associated to the atom mass in the calculation of the phase accumulation in the interferometer. Because of the large value of the Compton frequency compared to the frequency of microwave and optical atomic clocks (for example $\omega_C/2\pi \sim 3 \times 10^{25}$~Hz for a Cs atom), an improvement in precision by four orders of magnitude would result with respect to the best measurements of the gravitational redshift with clocks in space~\cite{Vessot1980}. Also, the measurement could be performed over extremely small distances ranging from micrometer to millimeter.
The concept was extended in \cite{Lan2013} proposing a Compton clock combining an atom interferometer with an optical frequency comb to link time to a particle's mass.
This interpretation aroused a controversy~\cite{Muller2010a, Wolf2010, Muller2010b, Hohensee2011, Wolf2011, Sinha2011, Hohensee2012, Wolf2012, Unnikrishnan2012, Schleich2013b}. 
%
This debate was useful not only to clarify the origin of the phase signal in an atom interferometer in the presence of gravity but also because it stimulated new ideas on possible experimental tests. 

The discovery potential of the analysis proposed in \cite{Muller2010a} was underlined in \cite{Hohensee2011} in the frame of the standard model extension \cite{Kostelecky2009}; it was shown that data from atom interferometers can be used to set stringent limits for equivalence principle violating terms.

As already mentioned in Sect.~\ref{sec:MeasG}, in \cite{Hohensee2012PRL} an experiment was proposed using atom interferometry to observe a gravitational analog of the Aharonov-Bohm effect \cite{Aharonov1959}.
Similarly to other topological phases induced by electromagnetic potentials that were measured also with atom interferometry (see for example \cite{Lepoutre2012} and references therein), in the case of the gravitational Aharonov-Bohm effect the phase shift is induced by the gravitational potential due to external source masses even if they do not produce a net classical force on the atoms. This experiments might be performed using  atom interferometry schemes with confined atoms as the ones demonstrated in \cite{Zhang2016,Xu2019}.

An experiment that would provide a test of the general relativistic notion of proper time in quantum mechanics was proposed in \cite{Zych2011} considering a Mach-Zehnder matter-wave interferometer in a homogeneous gravitational field. If the particle has an internal degree of freedom acting as a clock and if the two arms of the interferometer are separated along the direction of the field, according to general relativity and quantum complementarity the interference visibility will drop because proper time flows at different rates in different regions of space-time thus providing which-path information \cite{Zych2016}. The demonstrated quantum superposition at the metre scale \cite{Kovachy2015} combined with optical  clock
states could  enable the investigation of this effect in large-scale atom interferometers (Sect.~\ref{sec:largescale}).
An experiment along these lines was recently started in Florence planning to use optical clock transitions of Sr and Cd atoms \cite{Poli2018_ICAP,Tinsley2019_ECAMP}. 

The concepts of phase in matter-wave interferometers and proper time and the possibility to use atom interferometry to measure special-relativistic and general-relativistic time dilation effects were discussed  in \cite{Loriani2019,Ufrecht2020,Roura2020,Roura2020_arXiv}. Light-pulse atom interferometry configurations were proposed and analyzed in detail that would allow to detect such effects with feasible experiments.

In this frame, the experiment testing the equivalence principle for atoms in a superposition of internal states reported in \cite{Rosi2017} (see Sect.~\ref{sec:TestsWEP}) is of relevance; the interesting prospect of performing conceptually similar experiments with  energy gaps larger than the hyperfine splitting, as for example  narrow optical transitions in strontium, was mentioned in the paper.

The intriguing relation of gravity with entanglement was discussed in \cite{Marletto2017,Bose2017,Geiger2018,Altamirano2018,Margalit2020} also considering possible tests with matter wave interferometry.

In the debate about the phase in atom interferometers in the presence of gravity, the experiments with neutrons that first showed a gravitationally induced phase shift in a matter-wave interferometer \cite{Colella1975,Staudenmann1980,Werner1988},  for which related issues had been discussed \cite{Greenberger1983}, were considered as a reference.

Experiments with neutrons also enabled the first observation of gravitational quantum bound states for a particle above a horizontal mirror that with the Earth's gravitational field generates a confining potential well \cite{Luschikov1978,Nesvizhevsky2002}.
The resolution in the probing of the gravitational quantum states was improved using resonance spectroscopy techniques \cite{Abele2010,Jenke2011,Abele2012,Cronenberg2018}.
In addition to its intrinsic interest,  this effect  was proposed as a method to investigate gravity at very small distances, dark energy, and dark matter \cite{Jenke2014,Cronenberg2018}.
An interesting prospect is to perform similar experiments with atoms,
using quantum reflection from surfaces or atom optics tools \cite{Wallis1992},
with the advantage of a much larger flux of the atomic sources compared to neutron sources.  Ultracold light atoms, such as hydrogen, could be used to resolve the energy levels that for neutrons are of the order of peV with spatial separations of the order of micrometers. In the future, such experiments might be performed with antihydrogen atoms \cite{Voronin2011} and perhaps with exotic atoms.

%
\section{Testing quantum gravity models}
\label{sec:QG}
%

Quantum gravity is  the  research towards a theory merging quantum mechanics and general relativity. Different directions are followed such as string theory and loop quantum gravity. 
This might look like a purely formal effort since the two ranges are disconnected by several orders of magnitudes and relevant effects might play a role only in extreme conditions, like in black holes, at 
length scales of the order of the Planck length $ l_{\rm P}= \sqrt{\hbar G/c^3} \sim 10^{-35}$~m, 
time scales of the order of the Planck time $t_{\rm P}= l_{\rm P}/c = \sqrt{\hbar G/c^5} \sim 10^{-44}$~s, 
and energies of the order of Planck energy $E_{\rm P}= \hbar/t_{\rm P}=m_{\rm P}c^2=\sqrt{\hbar c^5/G} \sim 10^{28}$~eV,
that will hardly be directly accessible with lab experiments. 

According to some models, however, the quantum structure of spacetime might produce tiny effects that could be observable in high-precision low-energy experiments \cite{Amelino2013}.
Examples are the precision tests of the Einstein equivalence principle and the tests of Lorentz and CPT symmetries  \cite{Laemmerzahl2006}. Quantum gravity models predict deviations from these symmetries due to modifications of the  metric structure of spacetime although there are no estimates of the size of the effects at low energies.

The effects of spacetime fluctuations at small scales predicted by quantum gravity could produce violations of the equivalence principle,  modifications of the spreading of wave packets, and losses of quantum coherence; possible experiments with cold atom sensors were discussed in \cite{Simon2004,Goklu2008,Goklu2009,Goklu2011,Gao2017}. A problem for these experiments is how to discriminate the extremely small relevant effects from different  signals and from effects due to technical background noise.

A different approach was proposed in \cite{Amelino2009}:  the results of atom interferometry precise measurements of photon recoil were reinterpreted to constrain modifications of the energy-momentum dispersion relation which are expected in quantum gravity models. This analysis  in the small speed limit is analogous to the one performed in the relativistic regime using astrophysical data \cite{Amelino1998}.
Using the data available at the time from experiments measuring the photon recoil with a relative precision $\sim 10^{-8}$ to determine the value of the fine-structure constant \cite{Wicht2002,Gerginov2006}, in \cite{Amelino2009} it was shown that bounds could be set for the model parameters that for the leading correction were only one order of magnitude away from the Planck-scale level.
Following this approach, the same data were used in \cite{Arzano2010} to
set experimental bounds on deformations of the energy-momentum composition rule that appear in models of deformed Lorentz symmetry in some quantum gravity scenarios.
Data from photon recoil experiments were also analyzed in \cite{Gao2016} to constrain parameters in different models of generalized uncertainty principle \cite{Kempf1995,Ali2009,Maggiore1994}.
In view of recent experimental results on photon recoil with a relative precision of about $ 10^{-10}$ \cite{Parker2018,Morel2020} and plans to achieve $10^{-11}$ \cite{Mueller2020_Priv_Comm}, it can be anticipated that constraints  could be placed  at the Planck-scale level for the parameters of the model in  \cite{Amelino2009}. It should be noted, however, that the increase in precision for the measurement of photon recoil does not necessarily correspond to the same increase in the precision of the bounds for quantum gravity modifications of the energy-momentum dispersion relation \cite{Amelino2009,Mercati2010}. An optimization of the atom interferometry scheme for these experiments would be required.

%
\section{Search for dark energy }
\label{sec:DEAI}
%

Cosmological observations of the expansion of the universe \cite{Ade2016} can be interpreted assuming the presence of a so called dark energy that would account for about $70 \%$ of the universe energy density \cite{Copeland2006,Huterer2017}. 

In this section, the experiments using atom interferometry with the specific goal to investigate some form of dark energy are described.
Depending on the theoretical model, data from different experiments described
in other sections can be interpreted in terms of dark energy such as 
tests of the inverse-square law for gravity, 
limits in the parametrized post Newtonian metric, 
and tests of the equivalence principle 
\cite{Hui2009,Sakstein2018}.
%

The nature and properties of dark energy are not known. A possibility which is of relevance here is that it has the form of scalar fields; they should have an interaction with matter of the order of gravity but with a screening mechanism acting to suppress the effect of dark energy near dense materials in order to comply with experimental observations. Two such screened fields are the chameleon \cite{Khoury2004b} and the symmetron \cite{Hinterbichler2010}. 
The chameleon dark energy field can be characterized by two parameters, the first one associated to the self-interaction potential and the second appearing in the term for the interaction with ordinary matter. Symmetrons can be characterized by three parameters.

A possible scheme to search for dark energy in the form of a chameleon field using atom interferometry was proposed in \cite{Burrage2015} and the first  experimental results were reported in \cite{Hamilton2015}.
The basic idea is to place a small source mass inside the vacuum chamber where gravity is measured with the atom interferometer. Because of the screening, the scalar field is small at the chamber walls, rises to a maximum value inside the vacuum chamber and goes down near the source mass. Due to the gradient of the field, an atom is attracted towards the source mass with an acceleration that can be measured by atom interferometry. In order to discriminate the relevant effect from other effects, the source mass is moved from one position on one side of the atoms to another position on the other side.
 The scheme resembles the one used for the measurement of the gravitational constant $G$ with atoms \cite{Rosi2014} but in that case the bigger source mass was outside the vacuum chamber so that, similarly to Faraday shielding, the chameleon field inside would not be affected significantly \cite{Burrage2015}.
In \cite{Hamilton2015}, a Mach-Zehnder Raman interferometer in a vertical cavity with Cs atoms was used as the probe while the source mass was an aluminum sphere with a radius of 9.5~mm.  The results of the experiment were analyzed in detail in \cite{Elder2016}. In \cite{Jaffe2017}, thanks to several experimental upgrades, a sensitivity was obtained high enough to observe the gravitational attraction of the atoms by the source mass that was a centimetre-sized, 0.19 kg tungsten cylinder. The results, analyzed in terms of chameleons and symmetrons, led to an improvement by over two orders of magnitude on the limits  for the two models with respect to previous data. The results of this experiment were analyzed in terms of symmetrons in \cite{Chiow2020}. 

The results of an experiment conceptually similar to the ones in \cite{Hamilton2015,Jaffe2017} were reported in \cite{Sabulsky2019}. The main experimental differences were that Rb was used instead of Cs, no cavity was used for the interferometer light, the force was measured horizontally thus avoiding the large background due to gravity, and the source mass was a 19-mm-radius aluminum sphere. Results consistent with the ones in \cite{Jaffe2017} were obtained.

The prospect of performing these experiments in microgravity in order to increase the amount of time that the atoms spend near the source mass, thus allowing for greater sensitivity, was mentioned in \cite{Elder2016}. Experimental configurations for an experiment in the Cold Atom Laboratory (CAL) (see Sect.~\ref{sec:largescale}) on the ISS were discussed in \cite{Chiow2018}. Experiments could be performed also in drop towers \cite{Mueller2020_Priv_Comm}.

%
\section{Search for dark matter}
\label{sec:DMAI}
%

Different  astrophysical and cosmological observations (rotation curves of galaxies, gravitational lensing, cosmic microwave background) can be interpreted as the indication of the existence of what is called dark matter \cite{Bertone2018}. 
It would constitute about 27\% of the total mass-energy of the universe with the ordinary standard model matter being 5\% and  what is called dark energy 68\%. Dark matter would therefore make up about 84\% of the total matter in the universe.
However, we can say very little about its possible nature, properties,  mass, interactions. 

Most of the experiments trying to detect dark matter directly are based on particle physics methods; they search for heavy particles, with mass much larger than an eV, looking for energy deposition by dark matter particles in detectors   ~\cite{Feng2010,Akerib:2013tjd}. Most  of the efforts were on weakly-interacting massive particles
(WIMPs) with masses equivalent in the GeV-TeV range.
The lack of detected dark matter in the form of  particles led to alternative theories. 

Particle dark matter candidates were proposed which might have  masses smaller than an eV down to  $10^{-22} \, \mathrm{eV}$ and below~\cite{Marsh2016}. 
Examples are the pseudoscalar QCD axion and axion-like-particles and light scalar particles such as moduli, dilatons or the relaxion.
Such low-energy dark matter candidates would not be detected with traditional particle detection methods that are limited by their energy thresholds. 

New technologies are then required to search for such light dark matter candidates: atomic sensors can be relevant detectors in this range.

Existing models predict that possible effects of dark matter on standard model particles can be the precession of nuclear and electron spin, induced currents in electromagnetic systems,  acceleration of matter with violation of the equivalence principle, changes in the value of fundamental constants such as the fine structure constant and the electron mass \cite{Stadnik2015a,Stadnik2015b}. 
Detection schemes were considered based on models with dark matter in the form of  clumps or oscillating fields. 
Dark matter would then be detected as a transient effect if the atomic sensor crosses a dark matter clump~\cite{Pospelov2013,Derevianko2014,Stadnik2020} or as an oscillation effect at Compton frequencies for non-interacting fields~\cite{Arvanitaki2015}, even if stochastic~\cite{Derevianko2018}.
An overview of existing models and experimental tests can be found in \cite{Safronova2018}.

The experimental search for  axion and axion-like particles was described in \cite{Graham2015}. 
The results of experiments  using NMR spectroscopy to search for ultralight bosons such as axions, axion-like particles, or dark photons, were published in \cite{Abel2017,Garcon2019}.
%
%
Atomic magnetometers \cite{Budker2006a} have been used for the search of dark matter. Networks of such magnetometers \cite{Pospelov2013,Pustelny2013,JacksonKimball2018} can search for signals due to the coupling of dark matter  to atomic spins when the Earth goes through a dark matter compact object.

Different papers  were published in recent years using data from high precision atomic spectroscopy and atomic clocks to search for ultralight dark matter.
In~\cite{Tilburg2015}, spectroscopy of two isotopes of dysprosium was performed over a two-year span looking for coherent oscillations predicted if ultralight scalar dark matter with dilaton-like couplings to photons  induces oscillations in the fine-structure constant.
In \cite{Hees2016}, limits on possible oscillations of a linear combination of constants (fine structure, quark mass,  quantum chromodynamics mass scale) that would be produced by a massive scalar field were set using data from 6 years of accurate hyperfine frequency comparison of $^{87}$Rb and $^{133}$Cs atomic clocks. 
In \cite{Roberts2017}, 16 years of data from the GPS global positioning system were analyzed  to search for dark matter in the form of clumps. The motion of the Earth  through a galactic dark matter halo would perturb the GPS atomic clocks  due to the interaction with domain walls. Limits on  quadratic scalar couplings of ultralight dark matter to standard model particles were set.
In \cite{Kalaydzhyan2017}, possible effects of  dark matter  on the atomic clock stability were investigated. The dark matter was considered in the form of waves of ultralight scalar fields or as topological defects. The existing data for comparisons of ion clock frequencies allowed to set  limits on  dilaton dark matter. Prospects for experiments with microwave and optical clocks in space and with clocks based on nuclear transitions were discussed.
In \cite{Hees2018}, the consequences of a violation of the Einstein equivalence principle induced by light scalar dark matter were studied assuming models in which the  field couples linearly or quadratically to the standard model matter fields. Limits on the dark matter coupling parameters were obtained considering data from experiments testing the universality of free fall with masses made of different elements and from experiments comparing the frequency of different atomic transitions. The possibility of using atom interferometers was mentioned as an interesting prospect.
In \cite{Wcislo2018}, a network of Yb and Sr optical clocks operated in four laboratories in US, France, Poland, and Japan was used as an Earth-scale quantum sensor. The data analysis to search for topological defect and massive scalar field candidates was based on the different susceptibilities to the fine-structure constant between the atoms and the reference cavities \cite{Stadnik2015a,Wcislo2016}.
New bounds on the coupling of ultralight dark matter to standard model particles and fields in the mass range of $10^{-16} - 10^{-21}$~eV were set in \cite{Kennedy2020} by frequency comparisons between a strontium optical lattice clock, a cryogenic crystalline silicon cavity, and a hydrogen maser.
In \cite{Roberts2020}, data from a European network of fiber-linked optical atomic clocks was used searching for coherent variations in the recorded clock frequency comparisons across the network. Considering topological defect dark matter objects and quadratic scalar interactions with standard model particles, constraints were placed on the possible interactions of such defects with standard model particles.
With the steady improvement of optical clocks and of the methods to compare different frequency references, it can be anticipated that the sensitivity with which these effects can be probed will advance significantly. Also, particular transitions with an expected high sensitivity for searches for ultralight dark matter can be chosen, as the one proposed in \cite{Safronova2018_PRL}.

In \cite{Delaunay2017}, it was proposed that new interactions between the electron and the neutrons mediated by light new degrees of freedom can be probed by precision measurements of the isotope shift for two different clock transitions and four zero nuclear spin isotopes. 
The effect of the interactions would be detected as a deviation from the expected linearity in the so called King plots, that is, in the plot of the measured isotope shifts for one transitions vs the shift for the other.
Recent results of experiments were reported in \cite{Counts2020,Solaro2020}. 
In \cite{Solaro2020}, the isotope shifts for the 3d $^2$D$_{3/2}$ - 3d $^2$D$_{5/2}$ fine structure transition was measured for the five stable zero-spin isotopes of Ca$^+$ and combined with measured isotope shifts for the 4s $^2$S$_{1/2}$ - 3d $^2$D$_{5/2}$ transition. No nonlinearity was found in the King plots within the experimental uncertainties. In \cite{Counts2020}, instead, the measurement of the isotope shift for the  $^2$S$_{1/2}$ -  $^2$D$_{3/2}$ and $^2$S$_{1/2}$ - $^2$D$_{5/2}$ narrow optical transitions for five zero-spin isotopes of Yb$^+$ showed a $3 \times 10^{-7}$ deviation from linearity at the $3 \sigma$ uncertainty level. Further theoretical and experimental investigation is needed to ascertain the origin of the observed nonlinearity.
%
%



%


The possibility of using atom interferometry for the investigation of dark matter was  discussed in \cite{Arvanitaki2015}. The potential of the 10-m-scale atom interferometers
to constrain dilaton coupling parameters was analyzed in the frame of the model of Ref.~\cite{Damour2010}. In \cite{Graham2016}, the same experimental configuration was considered for different models of dark matter. As noted above, in \cite{Hees2018} the consequences of a violation of the Einstein equivalence principle induced by light scalar dark matter in certain models were investigated mentioning also the prospects of experiments based on atom interferometers on ground and in space, although the required level of precision has not been reached yet. 
 
In \cite{Arvanitaki2015}, the sensitivity to ultralight scalar dark matter waves of future large-scale optical and atomic gravitational wave detectors in space was  discussed. Other possible signatures of dark matter in optical interferometric gravitational wave detectors were discussed in \cite{Grote2019}.
The sensitivity to ultralight scalar dark matter of future atomic gravitational wave detectors on ground and in space was investigated in \cite{Arvanitaki2018} considering atom interferometers based on Sr atoms as proposed in \cite{Graham2013} and preliminarly demonstrated in \cite{Hu2017}. In particular, the advantage of such single-arm atomic detectors over optical interferometers for the detection of scalar dark matter was pointed out. 

In \cite{Graham2018}, a new atom interferometry scheme was proposed for the detection of axionlike ultralight dark matter. In this model, the axionlike particles act as a time-oscillating magnetic field coupling to spin thus inducing a phase shift as the atoms evolve due to the modified Hamiltonian. The achievable sensitivity was estimated considering an interferometer operated on the $^1$S$_0 - ^3$P$_0$ clock transition  of $^{87}$Sr which has a nuclear spin I=9/2.  

In \cite{Geraci2016}, it was proposed to use an apparatus based on two atom interferometers to search for dark matter composed of virialized ultralight fields considering time-varying phase signals induced by coherent oscillations of dark matter fields due to changes in the atom rest mass and changes in Earth's gravitational field. Configurations with a separation between the interferometers ranging from 1 km on the Earth to 1000-2000 km in space and interferometer sensitivity ranging from that expected to be achievable in the near term to a projected future prospect were considered. 

Data from a network of sensitive superconducting gravimeters were analyzed in \cite{McNally2020,Hu2020_EPJD,Horowitz2020} searching for time-dependent signals that would be produced when dark matter interacts with the Earth. In this frame, also atom interferometry gravimeters could be used \cite{Figueroa2020}.

In conclusion, experiments based on atom interferometry
can contribute to the search for dark matter. In particular, they are suited to look for ultralight candidates that cannot be detected using particle physics detectors.
Based on the existing proposals, it is clear that in order to search for dark matter, large-scale atom interferometry detectors are needed as for the detection of gravitational waves. Space detectors would significantly contribute to the search of dark matter by extending an Earth-based network of quantum sensors. The distance between the sensors in space and the Earth can allow to discriminate spurious signals. The past and ongoing activities for the development of large-scale atom interferometers and future prospects  are described  in Sect.~\ref{sec:largescale}.
%
%
The search for dark matter depends so much on the models, the parameter space is so large, and the required setups are usually so complex and expensive that experiments should be conceived and performed with apparatus developed also for other scientific goals as in MAGIS \cite{Graham2017,Adamson2018} and in the proposed SAGE and AEDGE space missions \cite{Tino2019,AbouEl-Neaj2020_AEDGE}. In these cases, indeed, the search for dark matter is a major objective; however, even in the case of a null result, the atom interferometers will enable other important experiments in gravitational physics with extremely high precision.

Finally, it is worth mentioning that models alternative to dark matter were considered to try and explain the observed dynamics of galaxies. A noticeable example is known as modified Newtonian dynamics (MOND)  where gravity changes at slow accelerations and galactic scales \cite{Milgrom1983a,Bekenstein2004}. Although it appears difficult to conceive experiments to test such models in lab-scale or near-Earth experiments, interpretations of precision gravity measurements from this point of view were proposed \cite{Gundlach2007,Klein2020}.

%
%
\section{Towards atomic gravitational wave detectors}
\label{sec:GWAI}
%
%

The interaction of gravitational waves with matter waves was early investigated theoretically in Ref.s~\cite{Linet1976,Stodolsky1979,Anandan1982,Borde1983} but the possibility of using atom interferometry to detect gravitational waves became an active field of theoretical and experimental research starting in the early 2000s. An overview of the initial ideas and efforts in this field can be found in \cite{Tino2011}. 

After a proposal for a compact atomic detector for gravitational waves \cite{Chiao2004}, that was shown to be flawed because of some mistakes \cite{Aspen2004,Roura2006,Tino2007},  configurations were proposed and analyzed for a single atom interferometer \cite{Aspen2004,Tino2007} and for two interferometers in a differential configuration also in combination with  optical atomic clocks \cite{Dimopoulos2008,Yu2011,Hogan2011,Graham2013,Hogan2016,Kolkowitz2016,Tino2019}. In both cases, the calculations showed that such detectors would typically require large-scale apparatus.  The differential configuration is presently considered as the most promising with a scheme using the ultra-narrow optical clock transition of Sr, or another alkali-earth or alkali-earth-like atom, as proposed in \cite{Yu2011,Graham2013} and preliminarly demonstrated in \cite{Hu2017}.

Ligo and Virgo optical interferometers are now well established observatories for  gravitational waves in a frequency range from about 15 Hz up to a few kHz. After the first detection in 2015 of the signal emitted by the coalescence of a pair of $36 M_\odot + 29 M_\odot$ black holes merging into a $62 M_\odot$ black hole \cite{Abbott2016}, signals  were detected from  other such binary black hole coalescences~\cite{Abbott2019a},  from low-mass compact binary inspiral \cite{Abbott2019b}, that  multimessenger data  showed to be produced by the merger of a binary neutron star system, and from the coalescence of a $23 M_\odot$ black hole with a $2.6 M_\odot$ compact object that would be the lightest black hole or the heaviest neutron star observed in a double compact object system \cite{Abbott2020}. The observed gravitational wave strain amplitude is of the order of $10^{-22}-10^{-21}$.
In the future, the proposed underground Einstein Telescope~\cite{Punturo2010_ET} would push the lower frequency limit down to $\sim 3$ Hz while the proposed space based LISA detector~\cite{Danzmann1996_LISA} would  enable the observation of low frequency gravitational waves, down to the mHz range, from very massive systems  and  from the fall of matter into supermassive black holes.

 Atom interferometers can indeed be designed to detect gravitational waves in the range of frequencies from a fraction of a Hz to a few Hz  that are lower than the ones accessible with present detectors and will not be accessible even if the future terrestrial and space large optical detectors  will be realized. 
In the case of a space detector (\cite{Tino2019} and references therein), atom interferometers represent an interesting alternative to optical interferometers with a potential simplification of the configuration; for example, a single arm would suffice instead of the double arm configuration required by optical detectors and the overall dimensions might be drastically reduced. 

The scientific motivation for the development of new detectors of gravitational waves based on atom interferometry is the prospect to observe sources that cannot be observed with other detectors  \cite{Rosi2018}. For example, they would enable the search for the merger phase of possible intermediate mass black holes, that is, systems of black holes with $\sim(10^3 M_\odot)$ mass. The detection of signals from such systems would demonstrate the existence of a ladder of black hole masses, from stellar mass to supermassive ones. The sensitivity needed to investigate, for example, a $10^3 M_\odot + 10^3 M_\odot$ binary black hole located at 3 Gpc  with SNR~$\simeq 5$ is about $10^{-21}$~Hz$^{-1/2}$ in the band $1 - 10$~Hz. 
In the  frequency range that would be covered by atom interferometers, also other effects could be investigated such as  type Ia supernova events~\cite{Seitenzahl2015} that are expected to emit neutrinos and  gravitational waves.



As mentioned above, calculations show that atomic detectors for gravitational waves should be km-scale in size for terrestrial apparatus and have a much longer baseline of thousands of kms and more for apparatus in space. 
The prospects for such large-scale atom interferometry apparatus and the work in progress are described in Sect.~\ref{sec:largescale}. 


%
\section{From lab-scale to large-scale atom interferometers on ground and in space \label{sec:largescale}}
%

As already mentioned,  different scientific applications require large-scale atom interferometry apparatus  either on Earth or in space. In this section,  proposed and ongoing activities in this direction are described. By large-scale, here we mean apparatus with a baseline of the order of 100 m and beyond but it is worth noting that the demonstration of 10-m size interferometers was instrumental to show the feasibility of larger apparatus.

The MIGA (Matterwave laser Interferometric Gravitation Antenna)
apparatus~\cite{Canuel2017} in Rustrel, France is presently under construction. It is based on 150-m-long horizontal optical cavities with an array of Rb atom interferometers along the  optical link to mitigate Newtonian noise~\cite{Chaibi2016}.
 The MAGIS (Mid band Atomic Gravitational wave Interferometric Sensor) project~\cite{Graham2017,Adamson2018} in US plans to develop a series of Sr interferometers  with increasing baselines of $\sim10$~m, $\sim100$~m, and $\sim1$~km. The 10-m baseline prototype is under construction at Stanford; 
the second, MAGIS-100, will be built at Fermilab in a 100-meter vertical  shaft at the NuMI neutrino beam facility. One atomic cloud will be located at the top of the shaft and one midway down thus allowing for $\sim 3$~s of free-fall and hence measurements at frequencies $<1$~Hz. 
The plan is to use $100-1000~\hbar k$ large-momentum-transfer atom optics and a cold atom flux of $10^6-10^8/$s.
The third one would be built in a km-scale vertical shaft at the Sanford Underground Research Facility (SURF).
The ZAIGA (Zhaoshan long-baseline Atom Interferometer Gravitation Antenna) is an underground atom interferometry  facility 
under construction near Wuhan, China. The design for the final apparatus includes a horizontal equilateral triangle configuration with two atom interferometers separated by 1 km in each arm, a 300-meter vertical shaft  with an atom fountain and atomic clocks,  1-km arm-length laser links between optical clocks~\cite{Zhan2019}. 

Other large-scale terrestrial apparatus were proposed. 
MAGIA-Advanced is an R\&D project for a large-scale atom interferometer based on ultracold rubidium and strontium atoms. The goal is to build an underground 100-500~m underground vertical apparatus in an existing shaft in Sardinia \cite{Rosi2018,Canuel2020}. In addition to the availability of such shafts from previous mines, the interest of Sardinia as a location for high sensitivity gravitational detectors stems from the extremely low seismic and anthropic noise. 
The AION (Atom Interferometric Observatory and Network) project in the UK is based on a proposal similar to MAGIS for a series of 
 $\sim 10$m, $\sim 100$m, and $\sim 1$km baseline atom interferometers ~\cite{Badurina2020}. The first stage would be located in Oxford; eventually, the full-scale detector would be networked with MAGIS.
ELGAR (European Laboratory for Gravitation and Atom-interferometric Research)  is a proposed European underground infrastructure with km-scale baseline \cite{Canuel2020}. It could include horizontal and vertical arms in the same location or in different sites.

The  terrestrial projects described above will be important to demonstrate the feasibility  of large-scale  atom interferometers and develop the necessary technology. 
Eventually, the km-scale apparatus could reach the sensitivity required for the detection of gravitational waves and search of dark matter, as described above.

The next frontier is to operate  cold atom sensors in space. 
The idea of experiments with cold atoms in space dates back to the early '90s with the first proposals and initial activities in France \cite{Lounis1993} and in Italy \cite{Albanese1992}.
As is usual for medium-scale and large-scale space missions, it took many years to develop the technology and to define the goals and the roadmap for gravitational physics tests using cold atom sensors in space.
Here, the main milestones and the first recent experimental demonstrations are summarized as well as future prospects.

The in-orbit operation  of an atomic clock based on cold rubidium atoms was first demonstrated in 2017 in the Chinese CACES (Chinese Atomic Clock Ensemble in Space) mission on board China's Tiangong-2 space laboratory~\cite{Liu2018}.
ACES (Atomic Clock Ensemble in Space) is an ESA project aiming to operating on the ISS an atomic clock based on cold Cs atoms~\cite{ACES}. It was the first cold atom space mission to be developed; the ACES payload is completing its qualification before the launch scheduled for mid-2021 \cite{Cacciapuoti2020}.
The planned ACES mission duration is 18 months, with the possibility of extending it up to 3 years. Microwave and optical links will enable space-to-ground clock comparisons that will be used to measure the clock gravitational redshift  with a target precision of  2 ppm. ACES will also search for time variations of fundamental constants by comparing ground clocks based on different atomic transitions.

The future prospect for cold atom clock experiments in space is to use optical clocks \cite{Poli2013,Ludlow2015} instead of the microwave clocks used in ACES and CACES. The SOC (Space Optical Clock) ESA R\&D activity \cite{Bongs2015} led to the demonstration of compact and transportable Sr optical clocks~\cite{Poli2014} and  the development of the relevant technology for a space mission on the ISS. Similar efforts are ongoing in China~\cite{Zhang2020}.

Atom interferometers would reach their ultimate performances in space.
As mentioned above, the phase accumulated in a Mach-Zehnder atom interferometer due to an acceleration 
depends on the square of the free evolution time $T$ between the  laser pulses of the interferometry sequence. 
On Earth, the maximum practical duration is $T\sim 1$~s with a free-fall distance of $\sim 5-10$~m. In space,  since the atoms and the apparatus are in free fall, an interrogation time  $T \sim 10$~s can be obtained
that on the ground would require an atomic fountain with hundreds of meters length. Therefore the interferometer sensitivity  can be increased by a factor $\sim 100$ or more in space with respect to a similar instrument on the ground with a much smaller size of the apparatus.
For some experiments, as for the tests of the equivalence principle, gravity gradients would still represent a limit but, as mentioned above, a method to compensate this effect has recently been proposed \cite{Roura2017} and experimentally demonstrated \cite{DAmico2017}.
The sensitivity of an atom interferometer operated as a gyroscope also increases with the free evolution time. 
Atom interferometers in space can also take advantage of a very quiet environment where vibrations, non-gravitational accelerations and other perturbations can be reduced to very low levels and the Newtonian noise is absent.

Several activities have been performed and are currently in progress to increase the so-called technology readiness level  and demonstrate the maturity of atom-based sensors for a space mission on the ISS and on satellites.
 
The HYPER mission proposal, submitted to ESA in 1999, was based on a Rb atom interferometer operated as an atomic accelerometer and gyroscope on a satellite orbiting around the Earth with the primary goal of a precise measurement of the Lense-Thirring gravitomagnetic frame-dragging effect \cite{Jentsch2004}. After an assessment and an industrial study, ESA decided not to continue the development of this mission because the technology readiness level was considered too low.

An R\&D activity for a  mission with atom interferometers in space was performed with the SAI (Space Atom Interferometer) project~\cite{Tino2007_space,Sorrentino2010_SAI,Sorrentino2011} funded by ESA.

Atom interferometry experiments with a Bose-Einstein condensate were performed in the Bremen drop tower \cite{Muntinga2013}. 
The $^{87}$Rb condensate was coherently split  and  the emerging wave packets were separated over macroscopic scales with the interferometer extending over more than half a second and covering distances of millimeters. Recently, evaporative cooling with an optical dipole trap in microgravity in the drop tower was demonstrated \cite{Vogt2020}. A similar experiment was performed using a smaller size elevator \cite{Condon2019}; all-optical production of a Bose-Einstein condensate was demonstrated.

A preliminary test of the weak equivalence principle for $^{87}$Rb and $^{39}$K was performed in parabolic flights~\cite{Geiger2011,Barrett2016}.  The experiments showed the possibility of reducing the effect of  common mode vibrations for a $^{87}$Rb-$^{39}$K differential interferometer \cite{Varoquaux2009, Barrett2015}. 

In 2017, a cold atom apparatus   was launched to a height of 243~km using a sounding rocket (MAIUS-1) \cite{Becker2018}. During the flight, several experiments were performed on laser cooling and trapping of atoms, observation of  Bose Einstein condensation, and study of the condensate collective oscillations under weightlessness conditions. Several building blocks were demonstrated for a future atom interferometry mission in space.

In 2018, NASA launched and installed on board the ISS the Cold Atom Lab (CAL).
CAL is a multi-user facility developed by JPL to investigate quantum gases in the microgravity conditions in space. It is designed to study ultracold and quantum degenerate samples of $^{87}$Rb, $^{39}$K, and $^{41}$K, including dual-species mixtures of Rb and K \cite{Elliott2018}. In \cite{Aveline2020}, the first scientific results  were reported showing free-space $^{87}$Rb BEC expansion times over one second in duration, and decompression-cooled condensates with sub-nK effective temperatures.
Prospects include the investigation of trap topologies enabled by microgravity, few-body physics, atom-lasers, and the test of techniques for atom interferometry in space. 
The follow-on of CAL is BECCAL, a collaboration of NASA and DLR. BECCAL
will also operate on the ISS with ultracold rubidium and potassium, different methods for coherent atom manipulation, and will offer new perspectives for experiments on atom optics  and atom interferometry \cite{Frye2019}. The possibility of an atom interferometry precision test of the weak equivalence principle with $^{85}$Rb and $^{87}$Rb atoms on the ISS was also discussed in \cite{Williams2016}.


A precision test of the equivalence principle on a dedicated satellite was the primary scientific goal of the proposed STE-QUEST (Space-Time Explorer and QUantum Equivalence Space Test) mission that was originally proposed  within the ESA Cosmic Vision programme  \cite{Aguilera2014,Altschul2015}. STE-QUEST was designed to test different aspects of general relativity: In addition to the test of the weak equivalence principle using atom interferometry, major objectives were a measurement of the gravitational redshift and tests of  Standard Model extensions. 
A study of STE-QUEST was performed in 2011 at the ESA Concurrent Design Facility at the European Space Research and Technology Centre with the identification of a preliminary design of the mission and its payload; it was followed by a 1-year industrial assessment study. In parallel, instrument studies were performed on design, interfaces, resources consolidation, performance budget analysis.
A  White Paper on STE-QUEST was submitted in 2019 to ESA in response to the call for ideas for  the Voyage 2050 Long-term planning of the ESA Science Programme \cite{Battelier2019}.



The SAGE (Space Atomic Gravity Explorer) mission proposal \cite{Tino2019} was submitted to ESA in 2016 in response to the Call for New Science Ideas in ESA's Science Programme. It is based on a multi-satellite configuration with payload/instruments including strontium optical atomic clocks, strontium atom interferometers, satellite-to satellite and satellite-to-Earth laser links. SAGE has the scientific objective to investigate gravitational waves, dark matter, and other fundamental aspects of gravity such as the weak equivalence principle as well as the connection between gravitational physics and quantum physics.
The AEDGE (Atomic Experiment for Dark Matter and Gravity Exploration) mission proposal \cite{AbouEl-Neaj2020_AEDGE} was submitted in 2019 to ESA in response to the call for ideas for  the Voyage 2050 Long-term planning of the ESA Science Programme. While keeping the same mission concept of SAGE, the potential for the investigation of dark matter was emphasized.
%

 As for other large-scale missions in space involving different satellites and a complex new technology, like for the proposed LISA mission, the approval and the development of the required complex technology can take decades;  based on space agencies strategies and available funding, earlier pathfinder smaller missions might be considered with reduced scientific goals that would allow to test the crucial technology. The  operation in space of a full-scale atomic gravitational observatory such as SAGE cannot be foreseen before 2040-2050.

%
\section{Conclusions} 
\label{sec:Concl}
%

Atom interferometers have been developed as new tools to investigate gravity. Their sensitivity, precision, range of applications are increasing steadily. They represent new instruments that we are using to look at nature. 

It is well known from the history of science that new instruments and more precise measurements often led to discoveries that stimulated a deeper understanding of fundamental physics.
Obviously we cannot anticipate whether  this will be the case for atom interferometry and gravitational physics but the ingredients are there: several aspects of gravity are presently not clear and the theory does not provide a complete description  of experimental observations; two theories, namely general relativity and quantum mechanics, provide a correct description  of different phenomena but are not consistent with each other; atom interferometers and clocks,  precise quantum sensors that were not available until recently, are scrutinizing new aspects of gravitational physics with increasing sensitivity. 

We are committed to take advantage of these new instruments and pursue the investigation of gravity with higher and higher precision. As described in this review, various are the paths presently investigated experimentally and planned for the future; possible unexpected results in one or more of these experiments might indeed point at new physics.\\

{\bf Acknowledgements}\\

The author acknowledges 
useful information and comments by Tim Kovachy, Holger M{\"u}ller, Franck Pereira dos Santos,  Yevgeny Stadnik, Jacques Vigu\'e, Nan Yu,
and the critical reading of a preliminary version of the manuscript by Dmitry Budker.\\

\bibliographystyle{unsrt}

{\bf Bibliography}\\



\end{document}